\documentclass{sig-alternate-10pt}
\usepackage{indentfirst}
\usepackage{cite}
\usepackage{amsmath}
\usepackage{amssymb}
\usepackage{dsfont}
\usepackage{graphicx}

\usepackage{algorithm}
\usepackage{algorithmic}

\usepackage{multirow}
\usepackage{flushend}
\usepackage{cuted}
\usepackage{setspace}
\usepackage{subfigure}
\usepackage{epstopdf}
\usepackage{enumitem}
\usepackage{booktabs} 
\usepackage{caption}
\usepackage[paperwidth=8.5in,paperheight=11in,columnsep=0.33in,lmargin=4.5pc,textheight=9.25in,textwidth=7in]{geometry}

\usepackage[draft,colorlinks=true,linkcolor=blue,citecolor=blue,urlcolor=blue,pdfauthor={Matthew Roughan},pdftitle={Rigorous statistical analysis of HTTPS reachability}]{hyperref}

\def\equationautorefname~#1\null{%
  (#1)\null
}

\newlength{\figwidth}
\newcommand{\eg}{{\em e.g., }}
\newcommand{\ie}{{\em i.e., }}
\newcommand{\etal}{{\em et al. }}

\newcommand{\Prob}[1]{\mathds{P} \! \left\{#1\right\}}

\begin{document}

\pagestyle{plain}

\title{Rigorous statistical analysis of HTTPS reachability}
\author{
  \begin{tabular}{ccc}
  George Michaelson\footnotemark[1]  
    & Matthew Roughan\footnotemark[2] \footnotemark[3] 
    & Jonathan Tuke\footnotemark[2] \footnotemark[3]  \\
  \email{\normalsize ggm@apnic.net} & \email{\normalsize matthew.roughan@adelaide.edu.au} & \email{\normalsize simon.tuke@adelaide.edu.au} \\
  \end{tabular} \\[6mm]
  \begin{tabular}{cc}
  Matt P. Wand\footnotemark[2] \footnotemark[4]  
    &  Randy Bush\footnotemark[5]   \\
  \email{\normalsize matt.wand@uts.edu.au} & \email{\normalsize randy@psg.com} \\
  \end{tabular}
}
\maketitle
 
 \begin{abstract}
   The use of secure connections using HTTPS as the default means,
   or even the only means, to connect to web servers is increasing.
   It is being pushed from both sides: from the bottom up by client
   distributions and plugins, and from the top down by organisations
   such as Google.  However, there are potential technical hurdles
   that might lock some clients out of the modern web. This paper
   seeks to measure and precisely quantify those hurdles in the
   wild.  More than three million measurements provide statistically
   significant evidence of degradation. We show this through a
   variety of statistical techniques. Various factors are shown to
   influence the problem, ranging from the client's browser, to the
   locale from which they connect.
\end{abstract}



 
\renewenvironment{itemize}{%
  \begin{list}{$\bullet$}{%
    \setlength\labelwidth{1.5em}%
    \setlength{\leftmargin}{1.5em}%
    \setlength{\topsep}{1pt}%
    \setlength{\itemsep}{-0.5mm}%
  }%
  }{\end{list}}

\newenvironment{ssitemize}{%
  \begin{list}{$-$}{%
    \setlength\labelwidth{2em}%
    \setlength{\leftmargin}{2em}%
    \setlength{\topsep}{-1mm}%
    \setlength{\itemsep}{-1mm}%
  }%
  }{\end{list}}

\newenvironment{sbitemize}{%
  \begin{list}{$\bullet$}{%
    \setlength\labelwidth{2.5em}%
    \setlength{\leftmargin}{1.1em}%
    \setlength{\topsep}{-1mm}%
    \setlength{\itemsep}{-0.5mm}%
  }%
  }{\end{list}}

\renewenvironment{enumerate}{%
   \begin{list}{\arabic{enumi}.}{%
    \setlength\labelwidth{1.5em}%
    \setlength\leftmargin{1.5em}%
    \setlength{\topsep}{4pt plus 2pt minus 2pt}%
    \setlength\itemsep{-0.6mm}%
    \usecounter{enumi}}%
  }{\end{list}}

\maketitle

\renewcommand{\thefootnote}{\fnsymbol{footnote}}
\footnotetext[1]{APNIC,  South
          Brisbane 4101, QLD, Australia.}        
\footnotetext[2]{ARC Centre of Excellence for
          Math. \& Stat. Frontiers.}
\footnotetext[3]{University of Adelaide,
          Adelaide 5005, SA, Australia}
\footnotetext[4]{University of Technology
          Sydney, Ultimo NSW 2007.}
\footnotetext[5]{Internet Initiative Japan (IIJ) Research, Tokyo, Japan.}
\renewcommand{\thefootnote}{\arabic{footnote}}

\providecommand{\tabularnewline}{\\}

\makeatother
 
\section{Introduction}

There is a growing push for ``HTTPS Everywhere,'' a self-explanatory
term for the ubiquitous use of HTTPS in preference to HTTP for all
services, not only those specifically requiring a secure
connection. The Electronic Frontier Foundation (EFF) is promulgating a
browser extension to this effect \cite{https_every} as a defence
against spying, \eg from nation states in the post-Snowden era. Google
supports the idea \cite{https_every_google}, and has announced that
they will give search-rank priority to HTTPS sites
\cite{16:google_https}.  And the increase in the number of
clients accessing the Internet through wireless connections mandates
encryption at the connection level. Reactions include the HTTPS-Only
Standard \cite{https_only_stand}, for the US Federal-Government.

%

HTTPS, or more exactly HTTP over TLS (Transport Layer Security) is a
secure form of the standard Hyper-Text Transfer Protocol. It is secure
in that it provides
\begin{enumerate}
\item server authentication using certificates, \ie a server can prove
  its identity;

\item a private communications channel, \ie it prevents eavesdropping;
  and

\item data integrity, \ie it prevents standard man-in-the-middle
  attacks. 

\end{enumerate}
HTTPS everywhere therefore appears beneficial.

There is a performance cost, however, documented
\cite{goldberg98:_compar_of_http_and_https_perfor,Coarfa:2006:PAT:1124153.1124155,Naylor:2014:CSH:2674005.2674991}
as far back as the 1990s. This cost arises primarily because the certificate
exchange requires an additional round trip at the start of a
connection. However, most HTTP requests don't require a full
handshake, and with modern hardware the cryptography overhead is not
critical. For example Doug Beaver from Facebook, stated ``We have deployed
TLS at a large scale using both hardware and software load
balancers. We have found that modern software-based TLS
implementations running on commodity CPUs are fast enough to handle
heavy HTTPS traffic load without needing to resort to dedicated
cryptographic hardware. We serve all of our [Facebook's] HTTPS traffic
using software running on commodity hardware.''
\cite{beaver12:_http2_expres_inter}.


So on the face of it, HTTPS Everywhere is a ``no brainer.'' There is
even an ``HTTP Shaming'' web page.


``HTTPS Everywhere'' seems to be happening. StatOperator \cite{https_usage} reported
that the number of (the top million) sites using HTTPS as the default
increased from around 103 to 116 thousand from July to September, 
2016. Google reports client usage statistics (via Chrome)
\cite{https_usage_google}, and they show similar steady growth from
2015 to the
present. Naylor~\etal\cite{Naylor:2014:CSH:2674005.2674991} reported
that in 2014, HTTPS accounted for 50\% of such connections.

However, there is an important question to answer before we convert
the entire Internet to HTTPS: {\em Will there be people who are stranded
behind port 80?}

We know that HTTPS is not an issue for many people (the current
large-scale deployments of HTTPS prove that it mostly works), but
there could be locations, or users of specific equipment that face
challenges. Detailed reasons are given in
\autoref{sec:background}. They range from concern about the
quality of the technology, to the rejection of compromised
connections.

In this paper, we provide evidence which will inform a
technical and policy debate about the deployment of secure web services in
the global Internet, by measuring whether users can access HTTPS in the
wild. We find that there {\em is} sufficient evidence to show
that HTTPS is {\em not} universally easily accessible.

A secondary concern of this paper is the statistical rigor necessary to
allow such a statement to be made with confidence. 
The proportion of users that
failed to make an HTTPS connection in our study was small. 
It has been common in the past to simply report
numbers, and to make bold statements unsupported by statistical
methodology confirming the conclusions. Here, we seek to present a
case study describing a standard (albeit often ignored)
statistical methodology for rigorously analyzing a set of measurements
to enable statistically confident statements to be made about the
results, despite a noisy and faint signal. The ability to detect
such faint signals is important --- a mere 1\% of users now represents
hundreds of millions of individuals.

Another advantage of this approach is that we can ask more
sophisticated questions. Google reports \cite{https_usage_google}
differences in the proportions of HTTPS access by device type and by region
for instance (with Japan notably lagging). The question is then, {\em
  is this difference significant, or simply ``noise''?} That is, can
the results be explained simply by natural statistical variation in
the sample, or are they real? We show here that the differences are real.

Between 10 November and 4 December 2016, we collected 3.3 million
observations using APNIC's web advertising infrastructure
\cite{michaelson2014experience}, which is similar in concept to other
cross-origin request measurements, \eg
\cite{Casado:2007:PTS:1973430.1973443,Burnett:2015:ELM:2785956.2787485}.
However, this measurement is different in two critical
respects. First, it uses Google's advertisement infrastructure to
redirect requests from a broader selection of end-users than might be
obtained by placing scripts in a limited number of
servers\footnote{\scriptsize Though the existence of widespread
  country-wide censoring will limit our visibility into certain
  domains.}. Second, the redirected requests are innocuous. Hence, in
this context, some of the potentially vexing ethical considerations of
such measurements are not at issue. The users are not redirected to
contentious third parties, and thus our experiments appear to the user
like any other advertisement to which users are subjected every day,
and which users can refuse to accept using common browser plugins (a
more detailed ethical discussion is included below).

We found statistically significant evidence that, in our sample, there
are clients that find HTTPS connections harder to complete than HTTP, 
and that this
difficulty was influenced (in order of importance) by origin
Autonomous System, browser, country of origin, and operating
system. The variety of effects on the results suggests that HTTPS
problems are caused by a range of problems, some of which are discussed
in \autoref{sec:background}.

Regarding the nature of the tests, and the rigorous statistical
analysis, we use two key ideas. First, the tests are constructed with
a control measurement, such that we can assess the natural level of
noise, and understand which connection failures might be statistical
variation, and which real.

Second, we apply a series on increasingly sophisticated statistical
techniques to investigate the data. Notably, we apply a technique
called {\em Rasch modeling}, which is used in assessing examinations
and other similar tasks, but whose use appears novel in the context of
problems such as Internet measurements. It is a tool that should be
more widely considered in the domain of analyzing Internet
measurements, but which required some degree of customization to our
types of problems. The set of tools applied illustrates the nature of
a statistical investigation, \ie it is not a matter of blindly
applying a test, but rather a process of investigation and refinement.

Finally, our anonymized dataset will be made public.


\section{Background and Related Work}
\label{sec:background}

\subsection{Experimental Context}

Simple web services with no protection against snooping or identity
are typically conducted over TCP port 80, using the HTTP protocol. We
call this `port 80' service or HTTP.

Web services which are protected by Transport Layer Security (TLS) are
usually conducted over TCP port 443, commonly called `port 443' or HTTPS. 

There have been many studies of HTTPS. However, they have focused on two main
topics:
\begin{enumerate}
\item the certificate landscape, \eg see
  \cite{Durumeric:2013:AHC:2504730.2504755,laurie12:_secur_inter,4768661},
  in which the problems with certificate distribution have led to
  security holes, and consequent fixes\footnote{\scriptsize TLS security is
    predicated on valid certificates, and there have been significant
    problems resulting from this weakness in the past. However, Certificate
    Transparency solves many of these issues
    \cite{laurie12:_secur_inter}.}; and

\item comparisons between HTTP and HTTPS performance, looking primarily at their latency difference, \eg
  see
  \cite{goldberg98:_compar_of_http_and_https_perfor,Coarfa:2006:PAT:1124153.1124155},
  but also considering communications overhead and
  energy consumption \cite{Naylor:2014:CSH:2674005.2674991}.

\end{enumerate}

As a consequence of using HTTPS, an additional handshake is needed to 
establish a connection.  There can be no effective proxy-caching of the content, and filtering (\eg by firewalls) is hampered. HTTPS also uses cryptography
which induces extra computational (and hence energy) costs, which may be
trivial on a modern computer, but may be important on battery-operated
devices, such as mobile phones. 

Additionally, a deeper consequence of the additional layer of complexity in
the operation of a simple request is the potential for
problems. Surprisingly, studies of HTTPS appear to assume basic
reachability, or more correctly, they appear to assume that HTTPS
reachability, while perhaps not perfect, will be no worse than 
HTTP.  However, it is not obvious that this will be so.  A
prominent browser maker asked if the Asia-Pacific Network Information
Centre (APNIC) Labs ad-based measurement system
\cite{michaelson2014experience} could see if a statistically
significant number of users were unable to access TLS protected web
resources.

So, what are the possible concerns? They range widely; the following
is an incomplete list.
\begin{enumerate}

\item A browser or OS may be too old to perform TLS at the current
  specification. The webserver used in this experiment  did not
  offer older approaches, such as RC4 cryptography, so there is a
  chance that pre-TLS 1.x browsers could not complete protocol
  binding. However, the older standards are no
  longer considered secure, and it is our view that providing a false 
  appearance of security is worse than providing none. 
  It might be tempting to tell users to ``catch up'',
  but this is a problem on mobile phone networks who sell captive
  locked phones left behind on ``old cold'' and deprecated protocol
  variants.

\item Some modern browsers use intermediate systems to speed up or
  cache data. Opera, for instance, deployed a worldwide ``anycast'' cloud
  of intermediates to offer speed-up services, performing tasks such as
  JPG compression, to make the web faster. It is possible that this service
  notionally works with TLS, but that it works badly for flows it has in port
  80 that move up into TLS because the state doesn't exist. Other
  well-meaning intermediary systems might break such up-lifts.

\item The additional overhead of the extra handshake makes the session
  more vulnerable to network problems, and hence less stable.

\item TLS protects against the threat of bad actor man-in-the-middle
  attacks.
  If an on-path attacker intercepts the session
  and attempts to hijack an aspect of the content, TLS will (and
  should) prevent the flawed connection. However, if such attacks are
  prevalent, they become DoS attacks on the HTTPS service.

\item A firewall along the path might block encrypted traffic as a
  matter of course.  Though most firewalls allow port 80 traffic,
  they sometimes block all other ports. This might be considered
  misconfiguration, but misconfiguration is not uncommon
  \cite{ranathunga2016a,wool2004}.

\item Flaws in implementations are possible, but configuration errors
  are more common \cite{ssl_tls_typic}.

\end{enumerate}

Our approach uses a cross-site reference within an advertisement in
order to create a measurement. The underlying idea is not new. It has been used to measure DNSSEC and IPv6
deployment, among other issues, \eg
\cite{Casado:2007:PTS:1973430.1973443} (or for a more general review
see \cite{Burnett:2015:ELM:2785956.2787485}). However, our approach
differs in several respects from 
\cite{Burnett:2015:ELM:2785956.2787485}. The most important is that
it performs a pair of measurements: a control based on HTTP, and the
actual measurement on HTTPS, the focus. As far as we
are aware, past studies have performed only the measurement of interest,
and therefore have been hard to interpret statistically.

However, APNIC's measurement infrastructure
also differs from other approaches in that we use
(paid) web-advertising to instantiate the tests (details described
below). Additionally, all fetches are to an APNIC-managed server,
avoiding the major ethical controversies of past experiments.

\subsection{Statistical Background} 

Here we lay out the key statistical background for the work to
follow, for readers who are not staticians. 
%

We start by defining basic terminology:

\noindent {\em Observation:} the collected responses of a single client's
  connection attempts.


\noindent {\em Measurement:} a particular feature of an observation, for
  instance, whether a successful HTTPS GET was completed. We also call
  these {\em response variables}, and denote them by random variables (RV)
  $Y^{(j)}$, where $j \in \{ \mbox{HTTP}, \mbox{HTTPS} \}$ is the measurement, and
  \[Y^{(j)} = \left\{
    \begin{array}{ll}
      1, & \mbox{ if measurement $j$ succeeds,}\\
      0, & \mbox{ otherwise.} 
    \end{array}
    \right.
  \]
  Observations are collected instances $\{ y_i^{(j)}\}_j$ of this RV.


\noindent {\em Test:} a statistical test applied to the data.

\noindent {\em Group:} a subset of the observations, divided by covariates, \eg by
  region, country, OS, browser, or ASN.

\noindent {\em Categorical variable:} a variable which takes a set of discrete
  labelled values. In our case, we have {\em nominal} categorical
  variables, \ie they are unordered. An example here is the type of
  browser being used (\eg Firefox v Chrome).

  \noindent {\em Predictor:} a variable, also called a {\em
    covariate}, whose value {\em may} influence the outcome of the
  measurements. Examples here include the country and Autonomous
  System (AS) of origin of the IP address.

The most important distinction here is between a {\em measurement},
and a statistical {\em test}.

Statistics has many tools, some aimed at data exploration or
visualization. However, one of the important ideas is the {\em
  hypothesis test}. We start with a well-posed pair of complementary
hypotheses: $H_0$ and $H_1$, with the former called the {\em null-hypothesis}. 
The statistical test will either
\begin{enumerate} 
\item reject the null-hypothesis, and thus substantiate the
  alternative $H_1$ with some degree of significance; or

\item fail to reject $H_0$, which is {\em not} the same
  as saying $H_0$ is true.

\end{enumerate}
In a single test there are two types of error:

\noindent {\em Type I:} We reject $H_0$ incorrectly, \ie when  $H_0$
is true.

\noindent {\em Type II:} We fail to reject $H_0$ when it is
false.


Any such test will be conducted with respect to a significance level,
$\alpha$, chosen at the outset of the experiment. Here we use the
commonly choice of $\alpha = 0.05$. This sets the Type I error
probability.


The procedure for the test is to calculate a test statistic, determine
from this a $p$-value, and then reject the null-hypothesis if the
$p$-value falls below the threshold $\alpha$. The common
interpretation of the $p$-value is that it is the probability, given
the null-hypothesis is true, of observing the given test statistic, or
a more extreme value.  Hence, a small $p$-value can be taken as
evidence that the null-hypothesis is invalid. However, we must be
careful of this interpretation, because of the underlying statistical
nature of the problem. Over a set of repeated experiments, even were
the null-hypothesis to be true, we would expect to see a uniform
distribution of $p$-values, including some values $< \alpha$
(Type I errors).

Given we reject the null, we can say we have statistically significant
evidence for the alternative hypothesis, at the prescribed level of
significance.

However, if we fail to reject the null, then we {\em must not} say the
null is true, as there are two possibilities: that the null is true,
or that we simply lack enough evidence to show that it is false. To
say the null is true would be to fall into the fallacy of ignorance,
\ie the fallacy that states ``because we have no evidence against H,
then H must be true.''

The Type I error probability is controlled by the parameter
$\alpha$. The Type II errors occur with a probability whose complement
is called the {\em power} of the test, and whose value depends on the
number of observations available, and on the data values
themselves. Thus we cannot control for it, but can ensure it is small
by providing enough observations.

This discussion applies to a single hypothesis test, for which we
consider {\em Per-comparison error rates} (PCER). However, as noted
above, if we perform multiple comparisons, we should expect some Type
I errors. Given a particular significance level, \eg $\alpha = 0.05$,
then we expect approximately $n \times 0.05$ such errors, and thus we
may draw the incorrect inference that a particular test is
significant. The overall probability of having at least one Type I
error in the family of tests is generally called the {\em Family-wise
  error rate} (FWER), and in multiple comparison tests, it is this
probability that we set. To do so, we must make a correction. A common
approach is the {\em Bonferroni correction}
\cite{shaffer95:_multip_hypot_test}, in which $\alpha$ is divided by
the number of tests in the family. We should note that this is rather
conservative, and that there are other more complex procedures
available \cite{shaffer95:_multip_hypot_test}, but our first goal here
is to move from a PCER, to a FWER. A key assumption underlying this
simple approach is that each test is independent of the others. A
discussion of how this relates to our measurements will follow,
although there are more advanced procedures to attack dependent
results, \eg \cite{westfall95:_multip_mcnem_tests}.

The advantage of a hypothesis test over visualization, or {\em ad hoc} 
conclusions, is that it is a consistent, repeatable test, with
strict, precisely defined assumptions and interpretation. Through
its use we can avoid making common errors, such as over-interpreting
limited evidence, or accepting logical fallacies.

There are also many criticisms of hypothesis testing. The most
important is, perhaps, that the value of a hypothesis test is entirely
dependent on the two hypotheses chosen. There are two tests that we
shall consider here. The first considers whether two measurements are
independent. The null-hypothesis therefore is that two random
variables are independent, and the alternative hypothesis is that they
have some dependence.

We will not derive the standard test statistics, and procedure for
calculating their $p$-values here. We simply note that there are
standard approximations that lead to a test within the class of
$\chi^2$ tests (see the Pearson and Spearman tests
\cite{agresti02:_categ_data_analy}), but that this test is simple
enough that an exact formulation can be derived, resulting in the {\em
Fisher exact test} \cite{agresti02:_categ_data_analy} for
independence. In our case, we will test that $(Y^{\scriptsize HTTP}, Y^{\scriptsize HTTPS})$
are independent random variables.  This is such a basic test that most
statistical packages contain a function for computing it. The main
limitation (leading to approximations) in the past was computational
cost, but this did not apply in our experiment so we used the
exact test.

We will discuss the interpretation of such results in our specific
context below, but one important limitation is that the measurements
may be {\em conditionally independent}, \ie they may appear dependent,
but actually they are only so because of an underlying dependence on
some latent or hidden variable. We will discuss this issue in more
detail below, by considering the obvious candidates for such a latent
variable.

However, our experiment is a {\em matched pairs} experiment. That is,
the pair of measurements is conducted on the same client. Thus we
should expect to see dependencies in the pair. The Fisher test is only
a confirmation of what we expect.

The real question of interest is whether some users have more trouble
with HTTPS than HTTP. This cannot be answered simply by 
comparing the proportions of successes for each measurement, because
these are matched pair experiments, and therefore very likely
correlated. Simply plotting the two probabilities, even with
confidence intervals, would not take these correlations into account.

However, the inclusion of our control experiment makes it possible to
ask this question in the formal context of hypothesis testing, using
{\em McNemar's test}~\cite{agresti02:_categ_data_analy}, another
standard component in the statistical toolbox. In this case, the
null- and alternative hypotheses are:
\begin{itemize}
\item $H_0$ is that $p_1 = p_2$; and 

\item $H_1$ is that $p_1 \neq p_2$; 

\end{itemize}
where $p_j$ is the probability that the $j$th measurement of any
particular observation is successfully completed. Rejecting the null
implies significant evidence that the difficulty of the two
measurements is different.

One criticism of hypothesis testing not discussed above is that it is
inadequate in itself. It can tell us ``if'' but not ``how much?''
There are many procedures and tools for taking the next step. 
These will be explored here, but since their use goes beyond simple
text-book applications, discussion is deferred until
\autoref{sec:analysis}.

\section{Experimental Method}
\label{sec:experiment}


\subsection{Measurements}

APNIC Labs uses web advertising to  measure browser behavior
worldwide \cite{michaelson2014experience}. The advertisement is written in HTML5 and fetches
multiple pixels in the various protocol exchanges under test (DNS, TCP/UDP, IP, TLS).
The system is 100 lines of JavaScript, gzip compressed to 5kb of data,
which is a small cost in web-page loading.

A {\em primer} query in each advertisement served causes the
user's browser to request a set of experiments with a unique client
identity (ID).

The primer query is an HTTP GET, and the body of the response is a set of
measurements to be performed. Each measurement is a discrete URL with the 
unique identity encoded, and is fetched under a ten-second timeout via asynchronous JavaScript web fetch; on completion of a measurement, the time is recorded.  
On completion of all measurements, or the ten-second timer, a {\em result}
web query is sent, which encodes the measurement
results in the query argument as a sequence of labels, showing the time or `null' if they did not complete inside the time limit. 

The web logs show whether a primer/result pair was valid, and if so,
we analyze the results. Observations without primer and result success
are filtered from the sample, although they can be useful
for later debugging and analysis.

The primary goal of these measurements was to collect information
about user web sessions and their ability to perform HTTPS. To do so,
the experiment tested whether a connection initiated on
HTTP could be taken to a TLS-protected session (HTTPS, port 443).
Google requires that advertisements placed over a
TLS-secured session remain in TLS. Thus we could not recruit TLS users
into a test of insecure web access. However, we were permitted to take
an unprotected port-80 HTTP session and include fetches of web
elements over TLS on port 443. Therefore, we could construct a
measurement where the port 80 users were asked to fetch one web asset
over TLS thus detecting their ability or inability to upgrade to
TLS.

The data were collected between the 10th of November and 4th of
December, 2016. \autoref{tab:data} shows the total set of client IDs,
and the number of valid responses.  A large number of connection
attempts defaulted to initiating over HTTPS. \autoref{tab:data}
shows the decrease in the number of experiments as we progress through
HTTPS to only HTTP. The latter is a small minority, but it is these that
we focus on, for reasons discussed below.
 
We focused on connections initiated over HTTP because
this HTTP signal provides a ``control.'' The priming
process and the HTTP control measurement follow an identical connection path. 
Hence, if the observation is valid, the client has demonstrated the ability to perform an HTTP GET; therefore failures of subsequent HTTP GETs provide an indication of the ``noise'' in the system, \ie the baseline rate of random loss
against which we should measure HTTPS connection failures. 

\begin{table}[!t]
\centering 
\caption{\small Experiment duration, and number of observations. 
  Analysis focuses on the 3.3 million experiments initiated with HTTP (with a
  subsequent HTTPS GET).}
\label{tab:data} %
\vspace*{-1mm}
\small
\begin{tabular}{cccccc}
\toprule 
\multicolumn{1}{p{1.5cm}}{\centering \mbox{ }\\ Duration \\ (days)}
& \multicolumn{1}{p{1.5cm}}{\centering Unique \\ client IDs \\ (mil.)}
& \multicolumn{1}{p{1.4cm}}{\centering Valid \\ responses \\ (mil.)}
& \multicolumn{1}{p{1.2cm}}{\centering HTTPS \\ init. \\ (mil.)}
& \multicolumn{1}{p{1.2cm}}{\centering HTTP\\ init. \\ (mil.)}
\tabularnewline
\midrule 
25 & 192.5 & 132.4 & 129.1 & 3.3
\tabularnewline
\end{tabular}
\vspace*{-4mm}
\end{table}

Initial simple statistical analysis strongly suggested that a
weak signal existed, but at an intensity which could not be
easily measured. The situation is analogous to experiments conducted
on mice who are genetically modified to have cancer. We wished to
measure factors that affected a situation which had a small probability
(cancer, in the analogy, or HTTPS connection failures in our case), and
so we ``artificially inflated'' the probability of seeing the phenomena of
interest. In our case, we focused on observations where the initial
connection was HTTP, because these were the cases where failures
of HTTPS were most often expected to occur.

As noted, it is a standard statistical approach to collect data in this way,
but we must note that the observation is not representative of a
``typical'' Internet user. For instance, were we to measure a failure
rate of 1\% on these observations, this does not mean that the general
population has a 1\% failure rate. However, the questions 
of interest us here are not the absolute value of the failure rate, but
whether HTTPS is ``harder'' than HTTP, and what factors affect the failure rate.

More formally, the main goal was to measure success/failure for
sessions upgrading to TLS and to see if those sessions which could not
upgrade to TLS were still successful on port 80.  In other words:
{\em ``are there stranded users?''} and {\em ``what factors affect the likelihood
of being stranded?''}

Google's infrastructure should have reduced duplicates, but we also removed
obvious duplicates from the data. There is some complexity in this
process, resulting from apparent fetches from the same IP address,
which cannot be resolved, due to the potential presence of middle-boxes such as
Network Address Translators (NATs). We preserved entirely unique
requests for the primer, but removed additional fetches without a new
primer. As a result, and because of DHCP reallocating IP addresses
over the timespan of the measurements, we cannot claim that there are
no duplicated observations, but they should be minimal.



 
\subsection{Data Collected}

The experiment logged all of the web fetches, using domain names 
directed to APNIC-managed DNS and web servers. We also captured the packet flow to relevant services: port 80, port 443, ICMP, DNS, as well as any fragmented IP state.

The combination of web logs, DNS logs, and packet captures allowed us
to collate experiments by their IP address and identity in the DNS
name, and as presented to the web. Thus we were able to derive the exact
sequences of events in any observation. 

In the case of this experiment, the data was processed into the form
of a series of flags indicating (1) the success or failure of each
stage, and (2) whether the measurement succeeded within a timeout. The
delays were recorded in each case up to 120 seconds, but for our
purposes we recorded success if the measurement completed within a
timeout of 10 seconds.

Only those users who issued a priming query in the same calendar day as they
sent a result were considered. This limited the exposure to 
repeated fetches, and excluded users who had a strong signal of
failure to perform basic fetches (neither the primer nor result had
any specific qualities which made them candidates to fail fetch).

In the data analyzed, unique client IDs were assigned to anonymize the
data.  We used code which harvests system entropy and time, to obtain
probably unique (modulo birthday paradox) non-sorted 96-bit numbers.
We then mapped them into hex (see \cite{txid}, for the code that was
embedded in the NGINX \cite{nginx} webserver).

\subsection{Classification of Covariates}

The secondary goal here was to identify the qualities behind the
quantities: \ie can we understand these users in terms of browser
type, ISP, economy, or operating system, in order to identify specific
problem causes? In practice, this is important because the goal behind
APNIC's participation in such experiments goes beyond simply finding
problems. Ideally, the experiment should also help develop strategies
to re-mediate any problems found.

For the purposes of the analysis, a set of qualities were identified,
which we felt were simple, easy to reproduce by other people, and
classified the users into different groupings for analysis, to try to
understand which (if any) qualified the problem. These qualities were:
\begin{itemize}
\item country,
\item region (based on United Nations sub-regions \cite{m49}),
\item origin Autonomous System Number (ASN),
\item browser, and
\item Operating System (OS).
\end{itemize}

We used the daily BGP table collected at AS4608 to map IP addresses
to Origin-ASN. There are well-known problems in such a mapping.
However, those problems are most prevalent in infrastructure
addresses, and we measured ``eye-balls'' here. That is, the IP
addresses we saw were associated with end users, for which the AS
mapping tools are a reasonable approximation
\cite{rasti10:_eyebal_ases}.  

Likewise, mapping of eye-balls to geographic locations is more
accurate \cite{rasti10:_eyebal_ases} than mapping arbitrary IP addresses to geographic locations. In this paper, we used MaxMind~\cite{maxmind} data to 
geolocate the IP addresses, but only at the country/region levels,
and so expected a reasonably low error rate.

We also logged each client's user-agent string, which provides details
of the client's browser, OS, and device. To collect and parse the
information we used the Python {\em uabrowser} library~\cite{uabrowser}. It
is known that the user-agent string is spoofed in some cases, for
instance the ToR browser bundle does so by default (\eg it pretends to
be running on Windows, regardless of the underlying OS). However, there
is no easy way to avoid this problem at present, and it remains a
caveat on the browser- and OS-level results.

We also considered categorizing the client's device-type, but this was
too noisy to be useful at this stage, due to the large number of
uniquely identified device types by vendor and version-string.

For each of these categories, we collated them into a series of unique
values, and then used a one-way random relabeling to anonymize the
categories. There might be enough data to perform some act of
deanonymization, to obtain values for some categories, However, it is important
to note that this level of blinding was not intended for the protection
of individual privacy (already protected through the client
ID anonymization, and unlikely to be compromised by the
additional coarse-grained categorisations). Rather it was intended to
allow the statistical analysis to proceed, unbiased by preconceived
notions of the likely results.


\subsection{Ethical Concerns}

Since Google's advertising infrastructure was used, we reviewed
compliance with Google and APNIC lawyers, and complied with Google's
legal restrictions on the measurements. In particular, this excluded
use of Personally Identifying Information (PII), which in any case we
did not require or want. End-user IP addresses were only used for AS
and regional classification, and were then anonymized via a
one-way-mapping.

The nature of the measurement technique precluded voluntary
recruitment, as in many other Internet measurements.  However, we
strictly followed any suggestion that the users wished to opt out of
such studies. For instance, end users who had enabled `do not track',
who had disabled JavaScript, or who ran ad-blocking software were not
recruited.  These measurements were also less contentious than others
that have applied similar cross-origin requests \eg
\cite{Burnett:2015:ELM:2785956.2787485}.  Evidence suggests that the
rate of TLS in the public web is high (above 50\%
\cite{Naylor:2014:CSH:2674005.2674991}) and very likely significantly
higher given the age of that study, and that it has been rapidly
increasing in recent years \cite{https_usage,https_usage_google}.
Therefore, the simple presence of a request to fetch a web asset over
TLS does not represent a high-risk activity.

We acknowledge that there is potential for users to hit firewall or
access control list limits, as a result of this experiment, or
register in economies which apply strong border checks on the protocol
of use. However, TLS is used ubiquitously for banking, login, end-user
tracking by less responsible advertisers, ``bread crumbs'' and
web-site logistics, and the measurement site to which the
advertisement redirected requests is innocuous, belonging to a
regional address registry (APNIC), so we were not able to discern any
reasonable risk to participants from such a connection.

 
Each user was requested to fetch one such asset, and as far as
possible, no users were repeatedly asked to run the experiment. We did
not seek to re-measure the same user.
There was some small possibility of repeated measurements: we could not
constrain re-presentation of cached state, and some forms of HTML5
embedding do re-cycle, such as YouTube cached state. Google does not
carry forward the referring site, and APNIC's web log does not record where
the advertisement was placed, so it is impossible to be certain
that there were no repeats, but the number should be small. 

In this experiment, those researchers not employed by APNIC were
exposed only to anonymized data.

\section{Analysis}
\label{sec:analysis}

In this section, we discuss the results of the analyses.  We will
start with ``broad brush'' simple hypothesis tests, then focus on
those same tests, applied to country, region, ASN, OS and browser.
This will be followed by a more comprehensive Rasch model, which
analyzes the data as a whole, and finally a generalised linear mixed
model will be applied.

\subsection{Standard Statistical Tests}

\begin{table}[t]
  \centering
  \renewcommand{\arraystretch}{1.2}
  \caption{\label{tab:tests} \small Statistical tests applied to
    the whole dataset.  Note that very small
    $p$-values are reported via a bound.}
  \vspace*{-2mm}
  \small
  \begin{tabular}{r|lll}
    Test   & $p$-value &  Accept/Reject  \\
    \hline 
    Fisher   & $< 2 \times 10^{-16 }$ & Reject null \\
    McNemar  & $< 2 \times 10^{-16 }$ & Reject null \\ 
  \end{tabular}
  \vspace*{-5mm}
\end{table}

The starting point for this analysis is to apply the two types of test
described above: the Fisher Exact test of independence/dependence,
and McNemar's Test to examine the difference between the two
measurements.

We applied all tests with a significance level of $\alpha=0.05$,
applying the appropriate Bonferroni corrections when conducting a set
of multiple tests (\ie we used significance $\alpha/n$ for a family of
$n$ tests). Note that in some cases, \eg when we were testing against
ASN, $n$ was quite large, and so the actual threshold was very
small. Less conservative corrections exist (for instance the Sidak or
Holm-Bonferroni) but the results here are conclusive without needing
the extra power gained through these more accurate corrections. 

The results for the test applied to the whole dataset are shown in
\autoref{tab:tests}.  We first see that we must reject the idea
that the measurements are independent.  A simple explanation is
that if it is it difficult to complete an HTTP connection, it may
also be difficult to complete an HTTPS connection. This is expected
for a matched pair experiment, so the Fisher Exact test is really just a
confirmation.

It is important to note the difference between independent {\em
  observations} and independent {\em measurements}. It is reasonable,
as a first approximation, to assume that observations of different
users are independent. The Fisher Exact test above shows that a paired
set of measurements are not independent.

\begin{figure}[t]
  \centering
  \includegraphics[width=\figwidth]{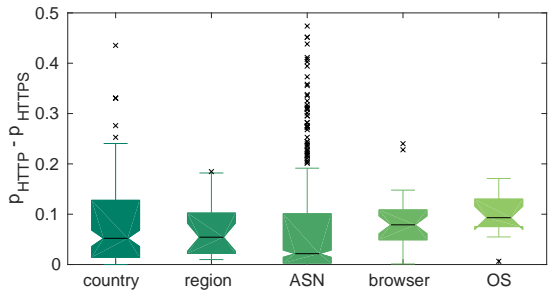}
  \vspace{-1mm}
  \caption{\small Boxplot of difference $p_{_{HTTP}} - p_{_{HTTPS}}$
    by covariate. Note that the lower quartile is always positive, as
    are all whiskers except those for ASN. If covariates had no effect
    on the result, these should all have similar interquartile range
    and median.}
  \label{fig:p-diff}
  \vspace{-4mm}
\end{figure}

The more subtle question that this raises is, given such
dependency, {\em can we really assert that HTTPS is ``harder'' than HTTP?}
The goal of assessing this question is addressed by McNemar's
test. The table shows that we must also reject the null in this case,
and hence we have significant evidence that the two measurements have
different levels of difficulty. The direction can be seen from the
differences in probabilities, \eg see \autoref{tab:p}, and \autoref{fig:p-diff}. 

This finding must be carefully qualified, noting here that although the
two measurements are matched, they typically occur in order, and
hence, there may be some effect on the second measurement resulting
from the state created by the first. So we must understand that this
experiment concerns lifting a connection up from HTTP to HTTPS, not an
arbitrary HTTPS connection (see the detailed notes in
\autoref{sec:experiment}).

\begin{table}[t]
  \centering
  \renewcommand{\arraystretch}{1.05}
  \caption{\label{tab:p} \small Number and proportion of correctly
    completed measurements by OS, and their difference.}
  \vspace*{-2mm}
    \small 
  \begin{tabular}{r|rrrrr}
             & & \multicolumn{2}{c}{Success proportion} &   \\
    OS ID    & $N/ 10^5$ & HTTP & HTTPS & Difference  \\
    \hline 
                   1  &  0.019 &  0.977 &  0.847 &  0.130   \\
                   2  &  0.077 &  0.983 &  0.917 &  0.066   \\
                   3  &  2.577 &  0.968 &  0.881 &  0.088   \\
                   4  &  0.655 &  0.978 &  0.903 &  0.075   \\
                   5  &  0.039 &  0.995 &  0.989 &  0.006   \\
                   6  &  7.337 &  0.969 &  0.881 &  0.088   \\
                   7  &  8.511 &  0.988 &  0.891 &  0.098   \\
                   8  &  0.036 &  0.978 &  0.873 &  0.105   \\
                   9  &  0.052 &  0.993 &  0.825 &  0.169   \\
                  10  &  0.647 &  0.981 &  0.881 &  0.100   \\
                  11  &  2.216 &  0.966 &  0.884 &  0.081   \\
                  12  &  0.155 &  0.974 &  0.820 &  0.154   \\
                  13  &  0.016 &  0.990 &  0.819 &  0.171   \\
                  14  & 10.566 &  0.963 &  0.909 &  0.055   \\
       \hline
    Overall  & 32.903 & 0.973 & 0.893 & 0.080 & \\
  \end{tabular}
  \vspace*{-5mm}
\end{table}
 

As well, care must be taken that the results have not been created by some
underlying covariate.  As an initial window into this question,
consider the boxplot \cite{mcgill78:_variat_box_plots,wickham12} shown
in \autoref{fig:p-diff}. The plot shows (shaded) the interquartile
range, and as a notch the confidence interval for the estimate of the
median value, of the difference $p_{_{HTTP}} - p_{_{HTTPS}}$. If the
covariates were truly irrelevant, we would expect that interquartiles
and medians should be the same (within the ranges of natural variation
shown by the confidence in intervals in the case of the median), and
hence the chart provides evidence that the covariates are important.

Therefore we now consider what part the covariates (country, region, ASN,
OS and browser) play.  We chose, at least initially, to be
conservative by only analyzing groupings with at least 500
measurements. It is quite possible that smaller groupings would have been
amenable to analysis, but we had no need (here) to describe the
relationships between all of the rarer groupings, as our goal was to
ascertain whether the overall result was supported on a finer level of
granularity.

The number 500 was chosen through an initial exploratory analysis,
which noted that some of the probabilities in question were quite
close to 1, and hence the typical statistical rules of thumb required a
moderately large number of observations. As we did not know exactly what
these probabilities were {\em a priori}, we chose a conservative lower
bound. We also found, as \autoref{tab:tests_covariates} shows, that
excluding the groups with a small number of observations excluded a  
small percentage of the data.

As noted above, we worked with anonymized data, \eg region codes were
remapped. We were not aiming here for strict anonymization of all
details, but aimed to (i) prevent individual identification without
extraordinary efforts, and (ii) ensure that the statistical analyses
were not biased by preconceived notions about different Internet
regions. Limiting the data to groups with at least 500 measurements
also facilitated this blinding.

\begin{figure*}[t]
  \centering
  \subfigure[\small Fisher exact $p$-values.]{
    \label{fig:fisher} 
    \includegraphics[width=\figwidth]{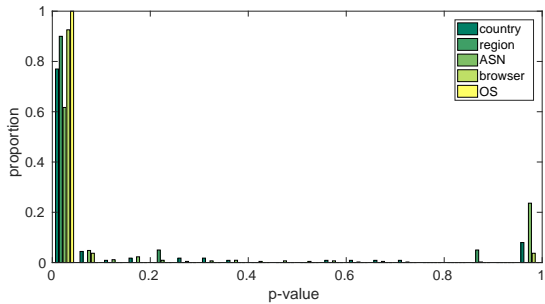}}
  \hspace{12mm} 
  \subfigure[\small McNemar $p$-values.]{
    \label{fig:mcnewmar} 
    \includegraphics[width=\figwidth]{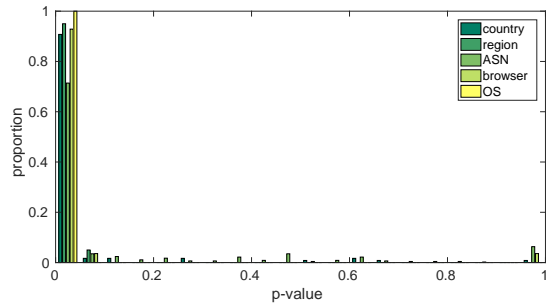}}
  \vspace{-4mm}
  \caption{\small The distributions of $p$-values for the Fisher Exact and
    McNemar tests, applied across country, region, ASN, browser and
    OS. The distributions in all cases vary dramatically from a uniform
    distribution, with values heavily skewed in the direction of
    $p=0$.}
  \label{fig:p-values}
  \vspace{-3mm}
\end{figure*}

The results can be seen in \autoref{fig:p-values}, which shows
histograms of the distributions of $p$-values over the set of tests
grouped by the various categorical covariates described above. The
important fact to note is that most of the $p$-values are small. We
cannot see (at the resolution of the plot) whether the $p$-values fall
below the threshold, so \autoref{tab:tests_covariates} summarizes the
tests, and shows that, in a large proportion of the cases, we rejected
the null-hypotheses. Thus we have significant evidence for these
properties in many of the groupings, and as such, our model should
incorporate these features.

Interestingly, ASN is the grouping with the lowest proportion of
rejected null-hypotheses, while we might have expected that ASN would have a
larger effect on the network aspects of the problem. However, remember
the large Bonferroni correction in this case, which leads to a rather
conservative test.

There is much more work to be done in this area. 
Multiple-comparisons could be applied, for instance, to differentiate
regions or some other covariates, in order to understand which
observed differences were significant, and which
were ``noise.'' However, in doing so there would be $O(n^2)$ comparisons
for $n$ regions, and more importantly, we are then performing a set
of hypothesis tests that are not all independent of each other,
complicating the test procedure greatly. Moreover, much of this
theory has been developed in domains where the conduct of each
measurement requires a physical or social experiment, and therefore
large sets of costly measurements. Instead, as our next step,
we opt to apply an approach called {\em Rasch modeling} described below.

\subsection{Rasch Modeling}

The disadvantage of the previous tests is firstly that they provide
only a yes/no answer (or really a yes/maybe answer), while we would like,
for instance, to be able to say whether a particular country is
falling behind in its ability to support HTTPS. Secondly, this
approach breaks the data into blocks and analyzes each separately,
but there are advantages in analyzing it {\em in toto}.

\begin{table}[t] 
   \centering 
   \renewcommand{\arraystretch}{1.05} 
   \caption{\small Hypothesis tests summaries for different covariates.
     $\tilde{N}$ is the number of groups left after exluding
     those with fewer than 500 observations. The \% of data is that retained by this filtering. And the two final columns
     report the proportion of tests for which we reject the null hypothesis over
     these groups for the Fisher and McNemar tests, respectively.}
  \label{tab:tests_covariates}
  \vspace*{-3mm}
    \small 
   \begin{tabular}{r|rrrr}
           covariate  & $\tilde{N}$  & \% of data  & Fisher  & McNemar \\
     \hline
             country  &  119 & 99.6 & 0.580 & 0.840 \\
              region  &   20 & 100.0 & 0.800 & 0.950 \\
                 ASN  &  458 & 93.1 & 0.277 & 0.555 \\
             browser  &   28 & 99.9 & 0.821 & 0.929 \\
                  OS  &   14 & 100.0 & 0.929 & 1.000 \\
   \end{tabular}
  \vspace*{-3mm}
\end{table}

There are many approaches that could be adopted to solve this problem. Here,
we consider an approach broadly called {\em Item Response Theory} (IRT).
This is an area best illustrated by its application to the analysis
of examinations. In this context, an exam consists of a list of $m$
questions, performed by $n$ students. At its simplest, each student
answers each question either correctly, or not, forming a set of binary
response variables $Y_i^{(j)}$, where $i$ is the student (in our context
an observation) and $j$ is the question (a measurement).

Within IRT, one of the oldest and most popular strategies is called
Rasch modeling
\cite{wright04:_introd_rasch_measur,wright77:_rasch}. This posits that
there are {\em latent} variables, namely:
\begin{enumerate}
\item the ability or proficiency of student $i$, denoted $\alpha_{i}$; and
\item the difficulty of question $j$, denoted $\beta_{j}$
\end{enumerate} 
that determine the probability of a student $i$ answering question $j$
correctly. The variables are latent in that we do not know them {\em a
 priori}. 

In its simplest case, \ie  a dichotomous response, we have response
variables $Y_i^{(j)}$, which are Bernoulli random variables indicating
a successful answer to a question, whose probabilities are modeled as
\begin{equation}
  \label{eq:prob}
  p_i^{(j)} = \Prob{Y_i^{(j)} = 1} = \frac{\exp\left( \alpha_{i} - \beta_{j} \right)}
                             {1 + \exp\left( \alpha_{i} - \beta_{j}
                               \right)}. 
\end{equation}
The probability function above is described as {\em logistic}, and has
an inverse called the {\em logit} function. Taking the logit we get
\begin{equation}
  \label{eq:prob_logit}
  \mbox{logit}(p_i^{(j)}) = 
  \log\left( p_i^{(j)} / (1 - p_i^{(j)}) \right)
    = \alpha_{i} - \beta_{j}.
\end{equation}
Thus we have a simple linear relationship at this point. 

Rasch modeling is one of the simplest, and yet most enduring approaches to
IRT~\cite{wright04:_introd_rasch_measur,wright77:_rasch}, because it
has a number of appealing properties:
\begin{itemize}
\item It simplifies the relationships so that reasonable estimates can
  be made, even though we have only one instance of each student
  attempting each question;

\item Unlike the conventional statistical paradigm, where parameters
  are fit to data, and accepted or rejected based on the accuracy of
  the fit, in Rasch modeling the objective is to obtain ``data'' that
  fit the model, \ie the latent predictor variables.

 
\item The Rasch model embodies the principle of invariant comparison,
  in which (broadly speaking) the effect on the outcome of a question
  is separated into the affect of the respondent, and the question's
  difficulty.


  
\end{itemize}
The model is not limited to modeling examinations, but can be applied
to any set of observations such as we have.  In the traditional
dichotomous Rasch analysis, we analyze each student who answers
questions correctly or incorrectly, but in our data, we have (i) a
large number of observations, well above that normally considered
in IRT, and (ii) we have no particular interest in the performance
of individuals.


To be precise, we could consider each observation as one
``student'' participating in the measurements, and analyze them all
separately, but this would be both banal (since
performance at this level of granularity is immaterial) and needlessly
computationally complex. In practice, we would like to group
measurements into meaningful partitions (a partition is a collection
of subsets that are disjoint and cover the original set).

However, when this is done, we depart from the pure simplicity of the
basic Rasch model. There are at least two alternative approaches; we
might think of these subsets as either being comprised of:
\begin{enumerate}

\item a group of similar students, who have an underlying property in
  common (usually we assume members of the group have proficiencies
  that are random variables with a common mean and variance); or 

\item a group of repeated measurements of a ``student'' who
  corresponds to the particular subset, and the responses are now
  Binomial random variables corresponding to the number of correct
  measurements within the subset.

\end{enumerate}
The two assumptions lead naturally to different solution algorithms.

The first model still follows \autoref{eq:prob_logit},
but now we assume that $\alpha_i \sim N(\lambda_k, \sigma^2)$, where
$\lambda_k$ is the group mean proficiency, and $\sigma^2$ is the
standard deviation within the group, parameters which we need to
estimate.

This approach is perhaps more advanced, in that exact results are
known, and there are off-the-shelf solvers using {\em Marginal Maximum
  Likelihood Estimation} (MMLE). We use
IRTm~\cite{braeken09:_inves,irtm}, a Matlab toolbox allowing quite
general models to be estimated. MMLE also has the disadvantage that we
assume a model for the distribution of the grouped student
proficiencies.


Moreover, in its model a categorical variable with $m$ categories is
deconstructed into $m$ binary variables, each an indicator for one
possible state of the original variable. This has the advantage of
simplifying the model, but at the expense of greatly increasing the
number of variables. For instance, in our filtered data, we have 119
countries, and so we must construct a covariate vector consisting of
119 binary elements.  The result is that the estimation procedure takes
more memory and time.

The results are illustrated below, in conjunction with those of the
second approach. 

The second approach takes a simpler model, that 
\begin{equation}
  \label{eq:rasch_2}
  \mbox{logit}(p_i^{(j)})
    = \alpha_{k} - \beta_{j}.
\end{equation}
for $i \in G_k$, where $G_k$ is the $k$th group.  We now only estimate
a group proficiency $\alpha_k$, not individual proficiencies. This has
the disadvantage that it might not be able to fit the data as
accurately, but it does free us from distributional assumptions.

However, the standard Binomial Rasch models for repeated measurements
assume each measurement is repeated a fixed number of times. For
instance, in partial-credit Rasch models
\cite{masters82:_rasch_model_partial}, a student may obtain some
proportion of the marks for a question, but each student answers the
same question, with the same total possible marks. But in our
groupings, the number of ``total marks'' would vary, depending on the
number of observations that fall into the group. This case does not appear
to have been treated in the literature, and hence we wrote our own
Alternating Least Squares (ALS) algorithm (also in Matlab) to estimate
the parameters. 

The algorithm alternates between fitting the $\alpha_k$ and the
$\beta_j$ values, keeping the alternative parameters fixed. It also
needs an additional fixed point of reference (because the variables
$\alpha_k$ and $\beta_j$ are not otherwise uniquely determined), which
we fix without loss of generality, to be
$E[\alpha_k]=0$. 


Ideally, we could group by all covariates at once. However, this
results in few measurements per group, and a {\em very} large number
of covariates in the MMLE approach, while in the ALS approach, we end up
with few observations in many of the bins, due to the combinatorial nature
of the number of bins. Thus we leave a combined covariate analysis
until later, and analyze each of the categories as separate groupings,
as we did in the statistical tests above. As before, we again made the
highly conservative choice to consider only groupings with at least
500 measurements, which is also helpful in refining the number of
categories for variables, and hence reducing the computation load for
MMLE.

\begin{figure*}[t]
  \centering  
  \subfigure[\small Computation times.]{
    \label{fig:rasch_times} 
    \includegraphics[width=\figwidth]{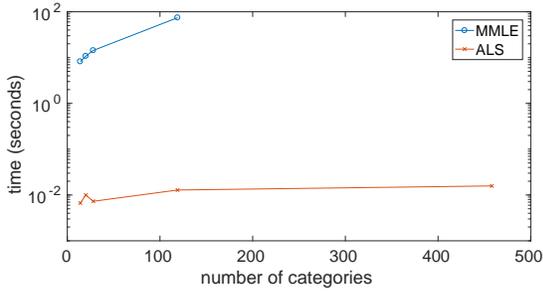}}
  \hspace{12mm}
  \subfigure[\small Errors.]{
    \label{fig:rasch_errors} 
    \includegraphics[width=\figwidth]{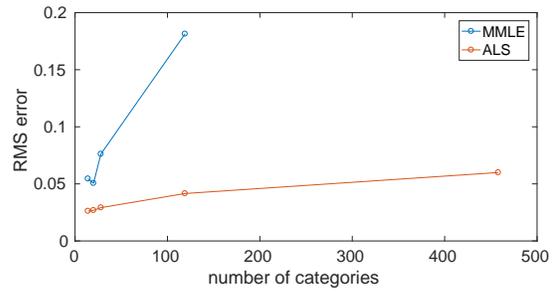}}
  \vspace{-4mm}
  \caption{\small RMS errors and computation times for the two approaches to
    Rasch modeling.}
  \label{fig:rasch_1}
  \vspace{-2mm}
\end{figure*}

We assessed the two approaches in this (somewhat non-standard)
application by comparing computation times, and Root-Mean-Square (RMS)
errors, as shown in \autoref{fig:rasch_1}. All computations were made
on an 8 core, Intel i7-6900K 3.2 GHz, running Linux Mint 18, and
Matlab R1016b. Note that the largest case for the MMLE algorithm did
not complete (within 24 hours), The ALS algorithm was orders of
magnitude more accurate and faster. 
So in what follows we focus on the ALS approach. Note also that the
estimation errors in ALS reach a maximum in the order of 5\%, which is
reasonable given the problem of interest, but it works better on the
cases with a smaller number of variates, and so we ignore ASN for the
moment.


The first detail to consider is the $\beta_j$ values, indicating the
difficulty in completing the two measurements (HTTP, and HTTPS). The
estimated values are shown in \autoref{tab:rasch_beta}, along with
their difference, which is an indication of the additional difficulty
of the HTTPS measurement. From all points of view, HTTPS is more
difficult than HTTP. The country/region and the browser/OS covariate
groupings also show some consistency, indicating that these variables
act on the measurements in somewhat the same way, which is not
surprising. 

\begin{table}[t] 
   \centering 
   \renewcommand{\arraystretch}{1.15} 
   \caption{\label{tab:rasch_beta}\small  ALS estimates of Rasch
     ``difficulty'' parameters with
     different groupings. Note the increase in difficulty for HTTPS.}
  \vspace*{-3mm}
    \small 
   \begin{tabular}{r|rrrr}
                      & country  & region  & browser  & OS \\ 
     \hline
           $\beta_{_{HTTP}} $  & -5.26 & -4.91 & -3.94 & -3.99 \\
           $\beta_{_{HTTPS}}$  & -2.92 & -2.74 & -2.29 & -2.12 \\
     \hline
          Difference  &  2.34 &  2.16 &  1.65 &  1.86 \\
   \end{tabular}
  \vspace*{-5mm}
\end{table}

\begin{figure}[t] 
  \centering
  \includegraphics[width=\figwidth]{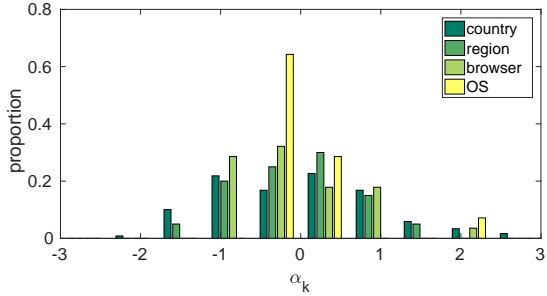}
 \vspace*{-3mm}
 \caption{\small Histograms of $\alpha_k$ values for ALS algorithm,
   omitting ASN, as this shows a less clear distribution.}
  \label{fig:rasch_alpha}
 \vspace*{-6mm}
\end{figure}

The second set of parameters to examine are the $\alpha_i$ values,
namely, the ability or proficiency of a particular covariate group to
perform the measurement, large values being
better. \autoref{fig:rasch_alpha} shows the distribution of $\alpha_i$
values for the region, OS, and browser covariates. We see that they
might be coarsely considered to follow a Normal distribution. The data
by OS fit this assumption least well, but remember that there are only
14 values here, so we do not expect to see a clean fit.  We expect to
see some natural variation here, because of estimation error, and
measurement noise. The excluded case, the data by ASN, present a much 
wider range of variation than the cases shown, and we hypothesize that
this variability is real, and important, but defer finalizing this
conclusion.

We also see outliers, here defined as those values that fall more than
1.96 times the standard deviation from the mean, \ie outside the 95th
percentiles. These are not extreme outliers, but these cases certainly
warrant further investigation. There are outliers on the positive
side: five countries, one browser, and one OS, each of which found the
measurements easier to correctly perform. These are of no particular
concern, though it might be interesting to investigate them further,
to understand if their success could be duplicated.  On the downside,
there is only one outlier country.

There are two main conclusions to be drawn. 
\begin{enumerate}
\item The measurements show that there is significantly more
  difficulty in performing HTTPS than HTTP measurements. The
  difference is often small, necessitating some extra care in order to
  determine whether the difference is significant.

\item There are country, OS, and browser differences, mainly important
  through a small set that exhibits more extreme variations from the norm.  

\end{enumerate}

Regarding the latter conclusion, we have deliberately kept the
blinding in our analysis here. Even though it is not necessary, and for
operational tasks such as improving the current situation, an
organization could use the same analysis to focus efforts on
improvements.  It is not within the ambit of this paper to make statements for or against particular regions, browsers or OSes.


This analysis has one major limitation --- we considered the different
groupings separately, instead of together. Moreover, we wish to
perform our investigations of outliers in more detail (particularly
for ASN), and explore how components interact, \eg in the above
analysis we tended to highlight countries that performed badly in both
tests, not just HTTPS.


\subsection{Generalized Linear Mixed Models}

The above approach is appealing in its simplicity, but deeper analysis
is possible. For instance, analysis of ASN proved problematic in the above
tests and analysis, at least in part because there are many more ASNs
than any of the other categories. Another statistical tool is a {\em
  Generalized Linear Model}
(GLM)~\cite{mccullagh89:_gener_linear_model}. This is a natural
generalization of linear modeling to cope with non-Gaussian random
variables (in our case, we have Bernoulli (0/1) random variables).

This has the potential to be applied to a wide variety of analyses,
but it requires refinement to the specifics of the problem of
interest, and we must also be able to perform the computations, which
will be more challenging than those of the previous tests.

In some ways, the idea of a GLM is similar to that of Rasch. We model
the probability of success via an inverse link function to a linear
combination of predictors. For Bernoulli data, we would (as in the
Rasch model) use the logit link function. However, there is a subtle
difference in the sense that in Rasch modeling, we think of each
student having a probability of success. Instead, we now think of a
generic Bernoulli response variable $Y^{(j)} \sim Bernoulli(p^{(j)})$,
for which we have observations, but now the probability of success is
a function of some set of predictors, \ie
\begin{equation}
  \label{eq:glm}
  \mbox{logit}\big( p^{(j)} \big) = 
  \beta_0 + \sum_{k=1}^{K} \beta_k X^{(k)},
\end{equation}
where $Y^{(j)}$ is the response we are interested in --- here it will
be $Y^{HTTPS}$, and the $\beta_k$ are $K$ parameters to be estimated,
linking the outcome to the predictor as a so-called {\em fixed effect}.

The predictors $X^{(k)}$ could include the covariates already
discussed (country, region, ASN, browser and OS), but note that as in
the MMLE Rasch analysis, categorical predictors such as region, with
$n$ possible values, must be converted into a binary vector of length
$n$, indicating which category applies. This leads us to the problem
of estimating a potentially large number of parameters $\beta_k$.  The
predictors could also include our previous test result $Y^{HTTP}$.




At least initially, the particular effect of each
country is not material. 
Our initial concern is to remove confounding effects of the
country from the analysis, so that the questions of 
interest can be answered. 
If this is the case, we can instead include a covariate as a
{\em random effect}, leading to the idea of a Generalized Linear {\em
  Mixed} Model (GLMM), incorporating a mixture of fixed and random
effects. In this formulation, we might take random intercepts $u_m$
and then
\begin{equation}
  \label{eq:glmm} 
  \mbox{logit}\big( p^{(j)}_m \big) = \beta_0 + \sum_{k=1}^{K} \beta_k X^{(k)}
                     +  u_m,
\end{equation}
where we now estimate different probabilities for each country $m$,
with a random-effect $u_k$ (which is a change to the intercept). We
could add similar random effects for the other covariates.  The key
difference here between the fixed-effect parameters $\beta_k$ and the
random-effects parameters $u_m$ are that we assume little about the
fixed-effect parameters, whereas the $u_m$ are assumed to come from a
Normal distribution, \ie $u_m \sim N(0, \sigma^2)$, whose variance
$\sigma^2$ is a parameter to be estimated here. This assumption allows
for estimation despite the potentially large number of values $u_m$,
because the model needs only the overall variance estimate, not the
individual parameters. Or, to be more clear, if we were to make
predictions from this model, we would not retain the individual $u_m$
values, but only the variance $\sigma^2$.

Random effects are used particularly when we have a predictor or
covariate that is not repeatable. For instance, in examinations, we
would not test students using the same questions twice. Here, the
countries observed are formed through the sampling process, and so the
number of observations per country is not set by the experimenters. We
cannot repeat exactly the same experiment at that level, so it is
reasonable to include these covariates as random effects. However, we
might want to change that decision if we were asking a particular
question about the impact of a particular country on the results, when
it would be appropriate to include a predictor for that particular
country as a fixed effect.  For instance, we could ask the question
{\em ``Is country X being left behind, taking into account the effects of
all of the other covariates?''} Including the other covariates as
random effects avoids the problem that we might believe country X is
being left behind, even though the result arises because country X has
a preponderance of users on OS Y, which is the real cause.

Another advantage of random effects is that they lead to relatively more
efficient algorithms for estimation, because we only have to estimate
one parameter instead of one per category. Here, we focus on the
analysis using $Y^{HTTP}$ as the only fixed effect, and consider which
covariates are important to include in any model. That is, we use
the analysis technique to explore whether we should incorporate region, for
instance, into our model.


There is a close connection between Rasch models and GLMM
\cite{doran07:_estim_multil_rasch_model}. In fact, the GLMM might be
thought of as a type of generalized Rasch model where the groupings
considered above are all potentially considered at once as random
effects. 

Here we perform a Maximum Likelihood Estimation (MLE) of the parameters
of these models using R~\cite{team13:_r}, and the
lme4~\cite{bates15:_mle4} package.

We do not consider arbitrary problems combining all factors at once,
due to the long computational times required for these
cases. Moreover, some factors are ``nested'' inside others, \eg
countries in regions, and to a lesser extent, browsers in OS. This
creates additional complexity in modeling this structure in the
regression. We will start by considering the simple problems, including
one covariate as a random effect at a time. 

Note also that in this section of the analysis, computational
requirements limit us to considering an sub-sample of our data. The
algorithms involved are very much more computationally complex, so for
the moment we sub-sample 10\% (randomly without replacement) of our
original set. However, we no longer restrict our attention to
groupings with over 500 observations.

The particular model we consider first is 
\begin{equation}
  \label{eq:glmm_us} 
  \mbox{logit}\big( p_m^{HTTPS} \big) = \beta_0 + \beta_1 Y_i^{HTTP} + u_m
\end{equation}
where observation $i \in G_m$, \ie the observation falls into group
$m$ of a particular covariate, included through a random intercept
$u_k$, and the fixed effect is the influence that the 1st measurement
has on the 2nd (the result). Here $\beta_1$ captures the marginal
increase in the chance of success in the second measurement, given the
first.

Apart from considering the question of quantitatively inferring the
effect of the first measurement on the second (we already know from
our Fisher test that there is typically some effect, but not how
much), we will also use this approach to assess which covariates are
necessary in our modeling enterprise. We will do so using Information
Criteria (IC), specifically Akaike IC (AIC)~\cite{Akaike74AIC} and
Bayesian IC (BIC)~\cite{Schwarz78BIC}.

These criteria allow one to determine an appropriate trade-off between
the accuracy usually obtained by including additional parameters into
a model, against the costs of addition parameters, particular in the
reduction of generality of the model, \ie the dangers of over-fitting.
In general, models with a lower IC are preferred, though there is no
single perfect IC
\cite{tune16:_compar_infor_criter_traff_model_selec}, and so 
two are used here to confirm conclusions. 

We adopt a greedy approach to inclusion of covariates: we start by
considering the benefit of including a single covariate, and we choose
the one that decreases the IC by the largest amount. We then consider
the benefit of including a second covariate and so on.  An advantage
this grants is that we can update an existing model, adding in extra
covariates instead of starting from scratch, saving additional
computation time.

\begin{table}[t]
  \renewcommand{\arraystretch}{1.05}
  \caption{\small
    Model selection. Covariates included in a model are indicated by their
    first letter. Selections are indicated in
    bold. Note that times for more complex models are the time taken to
    update the model from the next simpler model.} 
  \label{tab:selection}
  \centering
  \small
  \vspace*{-1mm}
  \begin{tabular}{l|rrrrrr}
    Covariates & AIC    & BIC     & time (s) \\
    \hline
    -          & 212244.1 & 212265.5    &  0.7 \\
    \hline
    C          & 202988.5 & 203020.5    &  34.0 \\
    R          & 206266.5 & 206298.5    &  26.5 \\
    A & {\bf 191988.6} & {\bf 192020.7} &  90.1 \\
    B          & 210905.8 & 210937.8    &  28.5 \\
    O          & 211560.8 & 211592.9    &  28.7 \\
    \hline
    A,C        & 191657.5 & 191700.2      & 153.0  \\
    A,R        & 191986.5 & 192029.2      & 253.5  \\
    A,B & {\bf 190974.9} & {\bf 191017.6} & 166.1  \\
    A,O        & 191424.6 & 191467.3      & 174.7  \\
    \hline
    A,B,C & {\bf 190554.3} & {\bf 190607.7} & 493.2  \\
    A,B,R      & 190922.2 & 190975.6        & 246.4  \\
    A,B,O      & 190729.5 & 190782.9        & 304.2  \\
    \hline
    A,B,C,R    & 190553.5 & 190617.5       & 384.9  \\
    A,B,C,O & {\bf 190187.1} & {\bf 190251.2}    & 397.3  \\
    \hline
    A,B,C,O,R  & 190178.3 & {\color{red} 190253.0}  & 574.9  \\
    \hline
  \end{tabular}
  \vspace*{-4mm}
\end{table}

\autoref{tab:selection} shows the process, where we have first
computed the GLM with only the fixed effect, then updated this model
to include one other covariate. Of the possibilities, ASN realises the
largest reduction in both ICs, and so we add this into our model.

We then update by adding a second covariate, the winning case being
browser, \ie the case listed as model (A,B). In the following steps the
winners are country and OS. Note that the selected covariates (see
boldface entries) present a decreasing sequence of ICs.

However, when we consider including region, we see that the IC do not
both decrease (specifically BIC indicated in red increases), and hence
we will not add this addition covariate. This is slightly
conservative, but the marginal improvement is at best very small.

The conclusion is that the alternative covariates add little
additional information. This is somewhat obvious: once we know the
country, we know the region. However, the process above (i) confirms
that all of the other covariates should be included, and (ii) confirms
that we prefer country over region as a explanatory covariate.

Once we have a model, we can examine it in detail. For the final model
(A,B,C,O), we rerun the estimation procedure for the full dataset. The
total analysis took 1.5 hours.  \autoref{tab:glmm_pars} shows the
results.

The small $p$-values for the fixed effects indicate a high degree of
significance. As we expect, the result of the first measurement is
related to that of the second. The positive value of $\beta_1$
indicates that success in the first measurement improves our chance in
the second.

The $\sigma_k$ values show the variance of the random effect
parameters for each type of covariate.  As might be expected, they are
decreasing because the are listed in order of importance as measured
by the IC. 

Also noteworthy is the scale of the effects. The $\sigma_k$ give a
measure of the variability of the random effect associated with the
covariate. It could be considered against the scale of $\beta_1$,
which tells the marginal increase resulting from success in the first
measurement. We can see that the variability of effects due to ASN are
of a similar magnitude to $\beta_1$, and the affect due to country is
not far behind. Effects due to browser and OS are somewhat smaller.


\begin{table}[t]
  \renewcommand{\arraystretch}{1.05}
  \caption{\small 
    GLMM estimated parameters of the full dataset, for model (A,B,C,O)
    including random effects of ASN, browser, country and OS. Note
    that standard error and $p$-values are not calculated for
    the $\sigma$ values.}
  \label{tab:glmm_pars}
  \centering
  \vspace*{-1mm}
  \small
  \begin{tabular}{l|rrr}
    Parameter & value & std. error & $p$-value \\
    \hline
    $\beta_0$          & 2.7960 & 0.0214 & $< 2 \times 10^{-16 }$ \\
    $\beta_1$          & 1.7236 & 0.0068 & $< 2 \times 10^{-16 }$ \\ 
    \hline
    $\sigma_{ASN}$     & 2.0971 & - & - \\
    $\sigma_{browser}$  & 1.3177 & - & - \\
    $\sigma_{country}$  & 0.7944 & - & - \\
    $\sigma_{OS}$       & 0.4181 & - & - \\
  \end{tabular}
  \vspace*{-4mm}
\end{table}

In any statistical analysis, care must be exercised when asserting
causality. Just because two events are correlated, we cannot
assert causality.
However, the GLMM
presented has the advantage that it can provide {\em
  predictions}. That is, we could use this model to predict the
outcome of our experiment in the future, based on the various
covariates, and hence provide some insights through this means. 

One last important insight is that the dependence on the covariates
indicates that the results are not an artefact of APNIC's measurement
infrastructure, as that would remove dependency on point of origin
effects. 

GLMMs have great capabilities, and we have only scratched the surface
here. We could consider cross-variable effects, hierarchical models,
and so on, and use this to build more focused experiments to
understand, for instance, the details of how some HTTPS connections
fail, in order to improve the situation.


\section{Conclusion and Further Work}

Using a large set of measurements, and detailed statistical modeling
we have shown that a small cohort of users in the real world will be
adversely affected if HTTPS is adopted universally. That cohort is not
a large proportion of Internet users, but those users deserve our
attention: {\em What are we going to do, to provide them with a secure
  network?}

We have categorized measurements by country and region, their provider
(origin ASN), browser and operating system, and shown that all of
these factors affect a client's facility with HTTPS. The range of factors
points to a range of causes for the blockages. The browser/OS
combination suggests a technological problem, but the other covariates
suggest problems based in the network near the clients.
 
In the future, we plan to further investigate, and use the
details of the analysis to help focus efforts onto relevant
development to mitigate the problem. We also aim to repeat this study
to consider the rate of change in this cohort of users, to understand
if this problem is growing, or shrinking over time.

The use of careful statistical methods was vital in this study. The 
underlying signal is weak, and hence required ``amplification'' and
careful analysis so as to be able to make confident statements. 




\section*{Acknowledgements}

We would like to thank the Australian Research Council for funding
through the Centre of Excellence for Mathematical \& Statistical
Frontiers (ACEMS), and grant DP110103505.

The Javascript code used by APNIC originates in a library written by Emile Aben, RIPE-NCC.

APNIC Labs has received support and in-kind assistance from Google,
ICANN, RIPE-NCC and ISC in conducting its web experiments.

  
\clearpage
{\small
\bibliographystyle{IEEEtran}
\bibliography{ms}
}\par\leavevmode

\end{document}